\documentclass[twocolumn,prc,aps,showpacs,preprintnumbers,amsmath,amssymb]{revtex4}

\usepackage{graphicx}

 \def\Pom{{ I\!\!P}}
 \def\Reg{{ I\!\!R}}
 \def\gsim{\mathrel{\rlap{\lower4pt\hbox{\hskip1pt$\sim$}}
 \raise1pt\hbox{$>$}}}

 \newcommand\la{\langle}
 \newcommand\ra{\rangle}
 \newcommand\beq{\begin{equation}}
 
 \newcommand\eeq{\end{equation}}
 \newcommand\beqn{\begin{eqnarray}}
 \newcommand\eeqn{\end{eqnarray}}
\def\mb{\,\mbox{mb}}

\def\GeV{\,\mbox{GeV}}

\def\lsim{\mathrel{\rlap{\lower4pt\hbox{\hskip1pt$\sim$}}
    \raise1pt\hbox{$<$}}}         
\def\gsim{\mathrel{\rlap{\lower4pt\hbox{\hskip1pt$\sim$}}
    \raise1pt\hbox{$>$}}}         
\def\Re{\,\mbox{Re}\,}

\def\mb{\,\mbox{mb}}

\def\GeV{\,\mbox{GeV}}

\def\T{T^h_A(b)}
\def\s0{\sigma_0(s)}
\def\sq{\sigma_{\bar qq}}
\def\st{\sigma_{tot}^{pN}}
\def\sel{\sigma_{el}^{pN}}
\def\sinhad{\sigma_{in}^{pN}}

\def\sta{\sigma_{tot}^{pA}}
\def\sela{\sigma_{el}^{pA}}

\def\beq{\begin{equation}}
\def\eeq{\end{equation}}

\def\beqy{\begin{eqnarray}}
\def\eeqy{\end{eqnarray}}

\newcommand{\ber}{\begin{displaymath}}
\newcommand{\eer}{\end{displaymath}}
\newcommand{\bey}{\begin{eqnarray}}
\newcommand{\eey}{\end{eqnarray}}

\def\beq{\begin{equation}}
\def\eeq{\end{equation}}

\def\beqy{\begin{eqnarray}}
\def\eeqy{\end{eqnarray}}

\begin{document}
\date{today}

\title{\bf Diffraction on Nuclei: Effects of Nucleon Correlations}
 \author{M. Alvioli}
 \affiliation{104 Davey Lab., The Pennsylvania State University,
      University Park, PA 16803, USA}
\author{C. Ciofi degli Atti}
\affiliation{Department of Physics, University of Perugia\\and\\
      Istituto Nazionale di Fisica Nucleare, Sezione di Perugia\\
      Perugia, Via A. Pascoli, I-06123, Italy}
\author{B. Z. Kopeliovich}
\author{I. K. Potashnikova}
\author{Iv\'an Schmidt}
\affiliation{Departamento de F\'{\i}sica, Centro de Estudios
Subat\'omicos, Universidad T\'ecnica Federico Santa Mar\'{\i}a,
\\and\\
Centro Cient\'ifico-Tecnol\'ogico de Valpara\'iso\\
Casilla 110-V, Valpara\'iso, Chile}
\begin{abstract}
\noindent The cross sections for a variety of diffractive processes
in  proton-nucleus scattering, associated with large gaps in
rapidity, are calculated within an improved  Glauber-Gribov theory,
where the inelastic shadowing corrections are summed to all orders
 by employing the dipole representation. The effects
of nucleon correlations, leading to a modification of the nuclear
thickness function, are also taken into account.  Numerical
calculations are performed for the energies of the HERA-B
experiment, and the RHIC and LHC colliders, and for several nuclei.
It is found that whereas  the Gribov corrections generally make
nuclear matter more transparent,  nucleon correlations act in the
opposite direction and have important effects in various diffractive
processes.
\end{abstract}

\date{\today}


\maketitle



\section{Introduction}

In hadron-nucleus collisions at high energies nuclei act almost like
"black" absorbers.  Therefore the optical analogy should be relevant
and diffraction appears to be an important process. Experimentally
diffraction appears as large rapidity gap events, when the debris of
the projectile hadron and the nucleus occupy only small rapidity
intervals close to the rapidities of the colliding particles. The
optical analogy is employed by the Glauber theory \cite{glauber} of
hadron-nucleus interactions, which assumes additivity of the
scattering phases on different bound nucleons. This is a single
channel approximation assuming that absorption, i.e. inelastic
interactions, generates via the unitarity relation only elastic
scattering. In reality, diffractive excitations of hadrons
frequently happen, and the Glauber approach was generalized to a
multichannel case by Gribov \cite{gribov}. The corresponding
corrections to the Glauber approximation are known as inelastic
shadowing, or Gribov corrections. Unfortunately, the multi-channel
problem needs detailed experimental information, which is mostly
unknown. One has to know all diffractive amplitudes, diagonal and
off-diagonal, for different diffractive excitations of the hadron.
Even the lowest order correction contains an unknown attenuation
factor for an excited state propagating through the nucleus
\cite{kk}.

One can sum up the Gribov corrections to all orders by switching to
the interaction eigenstates  \cite{kl}, which were identified  in
\cite{zkl} as color dipoles, and where the dipole approach to high
energy collisions was proposed. This phenomenology needs lesser
input and the key ingredient, the dipole-nucleon cross section, is
flavor independent and can be
 studied in different processes.

This method can be applied also to lepton- or photon-nucleus collisions
 \cite{krt1,krt2,kss2}, where leptons and photons display hadronic properties.
A detailed study of the inelastic shadowing corrections to different
diffractive channels in proton-nucleus collisions was performed,
within the dipole approach, in \cite{mine,kps}.

Here we are going to enhance the accuracy of the calculations
presented in \cite{kps}, by improving the model for the nuclear wave
function. Namely, most of calculations for nuclear shadowing effects
have relied so far on a simplified model of an uncorrelated single
particle density distribution in the nucleus. This model in
particular ignores the well known experimental evidences  for the
existence of a strong repulsion core between nucleons. Such a
repulsion should lead to short-range $NN$ correlations in the
nuclear density function, which in turn should modify the effective
nuclear thickness function controlling diffractive processes.

The consideration  of possible effects from nucleon-nucleon (NN)
short range correlations (SRC) appears to be particularly
interesting, in view of   recent experimental data on lepton and
hadron scattering off nuclei at medium energy, which provided
quantitative evidence  on SRC and their possible effects on dense
hadronic matter \cite{science}. Moreover, a recent calculation of
the total neutron-nucleus cross section at Fermilab energies has
indeed shown relevant effects from SRC even at high energies
 \cite{totalnA}.

\section{Glauber formalism}
\label{glauber}

The key assumption of the Glauber model is that the hadron-nucleus partial elastic amplitude at impact parameter $b$ has the eikonal
form \cite{glauber},
 \beqy
 \Gamma^{pA}(\vec b;\{\vec l_j,z_j\}) =
1 - \prod_{k=1}^A\left[1-
 \Gamma^{pN}(\vec b-\vec l_k)\right]\ ,
 \label{10}
 \eeqy
 where $\{\vec l_j,z_j\}$ denote the coordinates of an $i$-th target
nucleon; $i\Gamma^{pN}$ is the elastic scattering amplitude on a nucleon
normalized as
 \beq
\st = 2\int d^2b\,\Re\Gamma^{pN}(b)\ .
\label{20}
 \eeq
Further, one should calculate the matrix element of the amplitude
(\ref{10}) with the nuclear wave function,
$\psi_0(\vec{r}_1,\vec{r}_2, \dots  ,\vec{r}_A)=
\psi_0(\{\vec{r}_j\})\equiv |0\rangle$. Here we introduce new notations,
\beq
G_A(\vec{b})\,=\,\la 0|\Gamma^{pA}(\vec b;\{\vec l_j,z_j\})|0\ra = 1\,-\,\la 0|\prod_{i=1}^A\,
G^{pN}(\vec{b}-\vec{l}_i)|0\ra
\label{GiA}
\eeq
where
\beq
G^{pN}(\vec{b}-\vec{l}_i)\,=\,1-
 \Gamma^{pN}(\vec b-\vec l_i)
 \label{Gi}
 \eeq
The main problem in evaluating nuclear effects is therefore the
choice of the nuclear wave function.
 $\psi_0(1, \dots ,A)$.

 \subsection{Single particle approximation for the nuclear wave function}

 The most popular model for the
square of the  nuclear wave function appearing in the Glauber formalism
is the approximation of single particle nuclear density \footnote{We
ignore the effect of motion of the center of gravity assuming the
nucleus to be sufficiently heavy.},
 \beq
\left|\,\psi_o(\vec{r}_1,...,\vec{r}_A)\,\right|^2\simeq
\prod_{j=1}^A\,\rho_A(\vec b_1,z_1)
\label{40}
\eeq
where
\beq
\rho_A(\vec b_1,z_1)=
\int\prod_{i=2}^A d^3r_i\,
|\Psi_A(\{\vec r_j\})|^2.
\label{40bis}
 \eeq
 Within such an approximation the matrix element between the nuclear ground states reads,
 \beqn
&& \left\la0\Bigl|\Gamma^{pA}(\vec b;\{\vec l_j,z_j\})
\Bigr|0\right\ra
\nonumber\\ &=&
1-\left[1-{1\over A}\int d^2l
\Gamma^{pN}(l)
\int\limits_{-\infty}^\infty dz
\rho_A(\vec b-\vec l,z)\right]^A.
\label{30}
 \eeqn
 Correspondingly, the total $pA$ cross section has the form,
\begin{widetext}
\beqn
\sta &=&
2\Re\int d^2b\,
\left\{1 -\left[1 -\frac{1}{A} \int d^2l\,\Gamma^{pN}(l)\,T_A(\vec b-\vec l)
\right]^A\right\}\approx \nonumber\\
&\approx& 2\int d^2b\,\times
\left\{1-\exp\left[-{1\over2}\,\st\,(1-i\alpha_{pN})\,T^h_A(b)\right]\right\}
\label{50}
\eeqn
 \end{widetext}
 where $\alpha_{pN}$ is the ratio of the real to imaginary parts
of the forward $pN$ elastic amplitude;
 \beq
 T^h_A(b)= \frac{2}{\st}\int d^2l\,
\Re\Gamma^{pN}(l)\,T_A(\vec b-\vec l)\ ;
\label{51}
 \eeq
 and
 \beq
T_A(b) = \int_{-\infty}^\infty dz\,\rho_A(b,z)\ ,
\label{52}
 \eeq
 is the nuclear thickness function.
 We use the Gaussian form of $\Gamma^{pN}(l)$,
 \beq
\Re \Gamma^{pN}(l) =
\frac{\st}{4\pi B_{el}^{pN}}\,
\exp\left(\frac{-l^2}{2B_{el}^{pN}}\right)\ .
\label{53}
 \eeq

 Notice that in Eq.~(\ref{50}) and in what follows we use the exponential approximation
 of large $A$ only to simplify and clarify the formulas. For numerical calculations
 throughout the paper we always rely on the exact expressions, such as the first part
 of Eq.~(\ref{50}).

  The Glauber approach is a single channel model, therefore it is unable
  to consider diffractive excitation of the proton. However, a part of diffractive
   excitation of the nucleus occurs without excitation of the bound
   nucleons, when
    the nucleus just breaks up into free nucleons and nuclear fragments.
    Such events, $pA\to pF$, are called quasielastic and can be calculated
     within the Glauber approximation. Summing up the final states of the
nucleus $|F\ra$, applying the condition of completeness, and extracting
the contribution of the ground state of the nucleus, one gets,

\begin{widetext}
 \beqn
\sigma^{pA}_{qel} &\equiv&
\sum\limits_F\,\sigma(pA\to pF)-\sigma^{pA}_{el}
=\nonumber\\
&=& \sum\limits_F
\int d^2b\ \left[\left\la0\left|
\Gamma^{pA}(b)\right|F\right\ra^\dagger
\left\la F\bigl|\Gamma^{pA}(b)\bigr|0\right\ra
-
 \left |\left\la 0\bigl|\Gamma^{pA}(b)
\bigr|0\right\ra\right|^2\right]\nonumber\\
&=&
\int d^2b\left[\left\la0\left|
\bigl|\Gamma^{pA}(b)\bigr|^2
\right|0\right\ra
- \bigl|\left\la 0\left|\Gamma^{pA}(b)
\right|0\right\ra\Bigr|^2\right].
\label{70}
 \eeqn
 \end{widetext}
In the first order in nuclear density the first term in this
expression,  $\left\la0\left| \bigl|\Gamma^{pA}(b)\bigr|^2
\right|0\right\ra$, contains, besides the usual linear term
Eq.~(\ref{51}),  the quadratic term $\int d^2s\, T_A(\vec b-\vec
l)\left[\Gamma^{pN}(l)\right]^2 = T^h_A(b)\sel$. Both terms together
result in the exponent $\sigma_{in}^{pN}T_A(b)$. Then the
quasielastic cross section gets the form,
\begin{widetext}
 \beqn
\sigma^{pA}_{qel}(pA\to pA^*) =
\int d^2b\, \left\{
\exp\left[-\sinhad T^h_A(b)\right]-
\exp\left[-\st T^h_A(b)\right]\right\}.
\label{90}
 \eeqn
 \end{widetext}

\subsection{Nucleon correlations}

\begin{figure}[htp]
 \includegraphics[width=9cm]{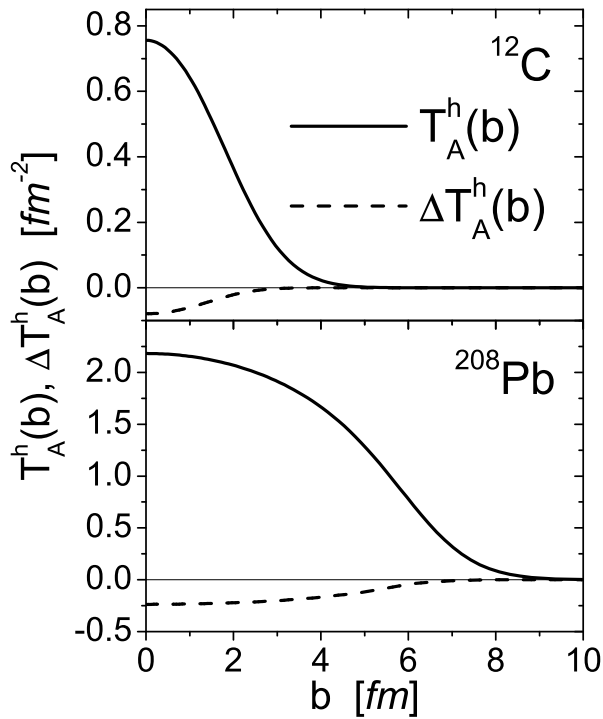}
 \caption{Nuclear thickness function $T_A^h(b)$ and the correction
 due to $NN$-correlations, $\Delta T_A^h(b)$, calculated at the energy of HERA-B, for carbon  and  lead, respectively. }
 \label{delta-T}
 \end{figure}

Equation (\ref{40}) represents only the lowest order term of the square of the full nuclear
wave function $|\psi_0|^2$. As a matter of fact, the latter can be written as an expansion
 in terms of
density matrices \cite{glauber,foldy} as follows:
\begin{widetext}
\beq
\left|\,\psi_o(\vec{r}_1,...,\vec{r}_A)\,\right|^2=\prod_{j=1}^A\,\rho_1(\vec{r}_j)
\,+\,\sum_{i<j}\,\Delta(\vec{r}_i,\vec{r}_j)\hspace{-0.1cm}\prod_{k\neq i,j}\rho_1(\vec{r}_k)\,+
\hspace{-0.5cm}\sum_{(i<j)\neq(k<l)}\hspace{-0.5cm}\Delta(\vec{r}_i,\vec{r}_j)
\,\Delta(\vec{r}_k,\vec{r}_l)\hspace{-0.3cm}\prod_{m\neq i,j,k,l}\hspace{-0.3cm}\rho_1(\vec{r}_m)
\,+\,\dots\,,
\label{psiquadro}
\eeq
\end{widetext}
in which the single particle density $\rho_{1}(\vec{r}_i)$ is
\beq
\rho_1(\vec{r}_1)= \int
\left|\psi_o(\vec r_{1},\vec r_{2},...,\vec{r}_A)\,\right|^2
\prod\displaylimits_{i=2}^A d^3r_i
\label{ro1}
\eeq
 and the  {\it two-body contraction} $\Delta$ is
\beq
 {\Delta(\vec{r}_i,\vec{r}_j)}\,=\,\rho_2(\vec{r}_i,
\vec{r}_j)\,-\,\rho_{1}(\vec{r}_i)\,\rho_{1}(\vec{r}_j)\,.
\label{contraction}
\eeq
The two-body density matrix
\beq
\rho_2(\vec{r}_1,\vec{r}_2)= \int \left|\psi_o(\vec
r_{1},\vec r_{2},...,\vec{r}_A)\,\right|^2
\prod\displaylimits_{i=3}^A d^3r_i
\label{twobody}
\eeq
satisfies the sequential condition,
\beq
\int\,d^3r_j\,\rho_2(\vec{r}_i,\vec{r}_j)\,=\,\rho_1(\vec{r}_i),
\label{sequential}
\eeq
which leads to the basic property of the two-body contraction
 \beq
 \int d^3r_j\,\Delta(\vec{r}_i,\vec{r}_j)\,=0.
 \label{deltazero}
 \eeq
Notice that the single particle density appearing in Eq.
(\ref{psiquadro}) is normalized to one, so that the densities
defined by Eq. (\ref{40bis}) and Eq. (\ref{ro1}) are simply related
by $\rho_A(\vec{r})\,=\,A\rho_1(\vec{r})$. It should be stressed
that in  Eq. (\ref{psiquadro}) only   unlinked contractions have to
be considered, and that the  higher order terms, not explicitly
displayed,  include unlinked products of 3, 4, {\it etc} two-body
contractions, representing contributions to two-nucleon
correlations, and  unlinked products of three-body, four-body, etc,
contractions, describing three-nucleon, four-nucleon, etc
correlations. We will give now a short  derivation of the total
cross section including two-nucleon correlations (more details will
be given elsewhere \cite{MCV}). Taking into account all terms of the
expansion (\ref{psiquadro}) containing all possible numbers of
unlinked two-body contractions, Eq. (\ref{GiA}) can be written in
the following form which yields the usual Glauber profile when
$\Delta = 0$ \cite{totalnA,MCV}:
\begin{widetext}
\beqy
\lefteqn{G_A({b})\equiv \int\prod_{k=1}^{A}d^3r_k
|\psi(1\dots A)|^2\prod_{i=1}^AG^{pN}(\vec{b}-\vec{l}_i)
\,= } \nonumber \\
& &=\,\int \prod _{k=1}^{A}d^3r_k\rho_1(\vec{r}_k) G^{pN}(\vec{b}- \vec{l}_k)
+ \sum_{i<j}\int \prod _{k=1}^{A}d^3r_k\Delta(\vec{r}_i,\vec{r}_j)
\prod _{l\neq i,j}^A
\rho_1(\vec{r}_l) G^{pN}(\vec{b}-\vec{l}_k) + {} \nonumber \\
& &+ \sum_{i<j\neq p<l}\int \prod _{k=1}^{A}d^3r_k
\Delta(\vec{r}_i,\vec{r}_j)\Delta(\vec{r}_p,\vec{r}_l)
\prod _{m\neq i,j,p,l}^A \rho_1(\vec{r}_m)G^{pN}(\vec{b}-\vec{l}_k)
+\cdots {} \nonumber \\
& & =G_A^{(0)}(\vec{b})+G_A^{(1)}(\vec{b})+G_A^{(2)}(\vec{b})+\cdots
\label{seriesGi}
\eeqy

\end{widetext}
where the superscript denotes the number of two-body contractions in the given term, each term
containing Glauber profiles to all orders. For each nucleus
we have considered all terms of the series (\ref{seriesGi}); the first term, corresponding
to the
single particle approximation of Eq. (\ref{40})
being
\beqy
G_A^{(0)}({b})\,&=&\,\int \prod _{k=1}^{A}d^3r_k\rho_1(\vec{r}_k) G^{pN}(\vec{b}-
\vec{l}_k) \nonumber\\
&=&\left[1- \frac{1}{A}\int d^3r_1\rho_A(\vec{r}_1) \Gamma^{pN}(\vec{b}-\vec{l}_1)\right]^A
\label{Gzero}
\eeqy
and the n-th terms
\beqy
G_A^{(n)}(b)\,=\,\frac{A!}{2^n\,n!(A-2n)!}\,X^n({b})\,Y^{A-2n}({b})
\label{Genne}
\eeqy
where
\beqy
X({b})=\int d^3r_1d^3r_2\Delta(\vec{r}_1, \vec{r}_2)
\Gamma^{pN}(\vec{b}-\vec{l}_1)\Gamma^{pN}(\vec{b}-\vec{l}_2)
\label{iks}
\eeqy
and
\beqy
Y({b})&=&\left[1- \frac{1}{A}\int d^3r_1\rho_A(\vec{r}_1)
\Gamma^{pN}(\vec{b}-\vec{b}_1)\right]
\eeqy
resulting from  the basic properties of the two-body contraction:
$\int d^3r_{i,j}\Delta(\vec{r}_i, \vec{r}_j)=0$ and
\begin{widetext}
\beqy
\int d^3r_1d^3r_2\Delta(\vec{r}_1,\vec{r}_2)G^{pN}(\vec{b}-\vec{l}_1)
G^{pN}(\vec{b}-\vec{l}_2)=\int d^3r_1d^3r_2\Delta(\vec{r}_1,\vec{r}_2)
\Gamma^{pN}(\vec{b}-\vec{l}_1)\Gamma^{pN}(\vec{b}-\vec{l}_2).
\eeqy
\end{widetext}
Eq. (\ref{seriesGi}) can now be written as follows
\begin{widetext}
\beqy
G_A({b})\,=
\,\sum_{n=0}^{A/2}\frac{A!\,X^n({b})[Y({b})]^{A-2n}}{2^n\,n!(A-2n)!}
\xrightarrow[A\gg 1]\,[Y({b})]^A\sum_{n=0}^{\infty}
\frac{A^{2n}X^n({b})}{2^n\,n!}\,=\,
[Y({b})]^A\,e^{\frac{A^2}{2}\,X({b})}\,.
\label{Gisum}
\eeqy
\end{widetext}
Using, for ease of presentation, the  optical limit approximation
\begin{widetext}
\beqy
[Y({b})]^A\,=\,\left [1-\frac{1}{A}\int d^3r_1\rho_1(\vec{r}_1)
\Gamma^{pN}(\vec{b}-\vec{l}_1)\right]^A\,=\,
e^{-\int d^3\,r_1\rho_A(\vec{r}_1)\Gamma(\vec{b}-\vec{l}_1)}{},
\label{Giexpo}
\eeqy
\end{widetext}
the insertion of  Eq. (\ref{Giexpo}) into Eq. (\ref{Gisum}) leads to the final
result:
\begin{widetext}
\beqy
G_A({b})&\simeq& 1-\exp\left[ -\int d^3r_1
      \,\rho_{A}(\vec{r}_1)\,\Gamma(\vec{b}-\vec{l}_1)+
\frac{1}{2}{\int d^3r_1 d^3r_2\,\Delta_A(\vec{r}_1,\vec{r}_2)\,
      \Gamma(\vec{b}-\vec{l}_1)\,\Gamma(\vec{b}-\vec{l}_2)}\right]=\nonumber\\
&=& 1-\exp\left[ -\frac{1}{2}\sigma_{tot}^{pN}\,\widetilde T_A^h(b) \right]
\label{finalGi}
\eeqy
\end{widetext}
where
\beq
  \Delta_A(\vec r_1,\vec r_2)=
  \rho_A^{(2)}(\vec r_1, \vec r_2)-
    \rho_A(\vec r_1)
    \rho_A(\vec r_2).
    \label{55b}
    \eeq
    which obviously differs by Eq. (\ref{contraction}) simply by a factor$A^2$, and
\beq
\widetilde T_A^h(b) =
 T_A^h(b) -\Delta T_A^h(b),
 \label{54}
 \eeq
 with  $T_A^h(b)$  given by Eq. (\ref{51}) and
 \beqn
  \Delta T_A^h(b)&=&
  \frac{(1-i\alpha_{pN})}{\sigma_{tot}^{pN}}
  \int d^2l_1\,d^2l_2\,
  \Delta^\perp_A(\vec l_1,\vec l_2)\nonumber\\ &\times&
  \Re \Gamma^{pN}(\vec b-\vec l_1)\,
  \Re \Gamma^{pN}(\vec b-\vec l_2),
  \label{55}
  \eeqn
  where
  \beq
  \Delta^\perp_A(\vec l_1,\vec l_2)=
  \int\limits_{-\infty}^\infty dz_1
  \int\limits_{-\infty}^\infty dz_2\,
   \Delta_A(\vec r_1,\vec r_2),
   \label{55a}
   \eeq
  is the transverse two-nucleon contraction.
  It can be seen that the inclusion of $NN$ correlations in nuclei leads
  to a modification
of the nuclear thickness function $T_A^h(b)\Rightarrow \widetilde T_A^h(b)$.
Due to
  its  general structure  and the
 basic property $\int d^3r_{1,2}\,\Delta_A(\vec r_1,\vec r_2)=0$, the sign
 of the
contraction  is mostly negative, with a small positive contribution at large
separations.
In Fig.~\ref{delta-T} we present  $T_A^h(b)$ and  $\Delta T_A^h(b)$ for
$^{12}C$ and $^{208}Pb$.
We see that $\Delta T_A^h(b)$ is indeed  mostly negative, so according
to the definition Eq.~(\ref{54}) correlations
increase the nuclear thickness function and make nuclear medium more opaque
\cite{totalnA}.
At the same time,  the corrections are small, $\Delta T_A^h(b)\ll T_A^h(b)$,
and the effects from higher order correlations, estimated in Ref. \cite{totalnA}, can safely be
disregarded.

 A short description of the way in which the one and two-body densities and
 contractions have been calculated is now in order. Following Ref. \cite{totalnA}.
 the two-body
density has been obtained from the  fully-correlated wave function of Ref.
\cite{alv01,alv02}, $\psi_0 ={\hat F} \phi_0$, where ${\hat F}
=\prod_{i<j}[\sum_{k=1}^8 f_k(r_{ij}){\hat O}_k(ij)]$ is a
correlation operator generated by the realistic Argonne
$V8^\prime$ interaction \cite{vuotto}, and $\phi_0$ a mean field shell model
wave function composed of Woods-Saxon single particle orbitals.
The above wave function largely differs from
the Jastrow one, featuring  only central correlations, since the
operator ${\hat F}$  generates central $({\hat O}_1=1)$, spin
$({\hat O}_2(ij)= {\vec \sigma}_i\cdot{\vec {\sigma}}_j)$, isospin
$({\hat O}_3(ij)= {\vec \tau}_i\cdot{\vec {\tau}}_j)$,
spin-isospin $({\hat O}_4(ij)= ({\vec \sigma}_i\cdot{\vec
{\sigma}}_j) \,{(\vec \tau}_i\cdot{\vec {\tau}}_j))$, tensor
$({\hat O}_5(ij)= {\vec S}_{ij})$, tensor-isospin $({\hat
O}_6(ij)= {\vec S}_{ij}\,({\vec \tau}_i\cdot{\vec {\tau}}_j))$,
etc. correlations. The two-body density and contraction therefore
reflect
not only the short range repulsion but also the spin-isospin dependence
of the interaction,
particularly that generated by the tensor force. The parameters of both the single particle wave
functions and the various correlation functions have been fixed from the ground-
 state energy calculation so that no free parameters are present in our approach.

The contraction $\Delta(\vec{r}_1,\vec{r}_2)$ resulting from our
calculation exactly satisfies the sum rule $\int d^3{r}_1
\Delta(\vec{r_1}, \vec{r}_2) = 0$, since the one-body density
$\rho_1(\vec {r}_1)$ exactly results from the integration of
$\rho_2(\vec {r}_1, \vec {r}_2)$. Notice, moreover,  that our
one-body point density and radii are in agreement with electron
scattering data \cite{electron}. We have also investigated the
validity of the approximation in which
 the nuclear matter
 two-body density
$\rho_2(\vec{r}_1,\vec{r}_2)\,=\,\rho_1(\vec{r}_1)\,\rho_1(\vec{r}_2)
\,g(|\vec{r}_1-\vec{r}_2|)$ is used for finite nuclei, finding that it leads
to a strong
violation of
the sequential relation
 $\int d^3 {r}
\rho_2(\vec{r}_1,\vec{r}_2)= \rho_1(\vec{r}_1)$ for nuclei with
$A < 208$. Thus, when  such an approximation is used  to introduce
correlations in light and  medium-weight nuclei,   a
mismatch between the one-body density (usually taken from the
experimental data) and the two-body density is generated.

Using the nuclear thickness function which includes the effects of correlations, Eq. (\ref{54}),
 the total  cross section, Eq.~(\ref{50}),  acquires a
correction, $ \sigma_{tot}^{pA} \Rightarrow \sigma_{tot}^{pA} +\Delta \sigma_{tot}^{pA}$, which is positive and
  can be approximated as,
  \beq
 \Delta \sigma_{tot}^{pA} \approx -
 \sigma_{tot}^{pN}\int d^2 b\,
 \Delta T_A^h(b)\exp\left[-{1\over2}\sigma_{tot}^{pN}T_A^h(b)\right].
 \label{55c}
 \eeq
 which is also positive, since $ \Delta T_A^h(b)$ is itself
 negative.

   We see that this correction to the total cross section comes mainly from
    peripheral collisions,
   and rises with $A$ rather slowly, as $A^{1/3}$.
 Notice that the accuracy of the optical (exponential)  approximation
in (\ref{50}) is quite good, $\sim 10^{-3}$ for heavy nuclei, but it
gets worse with decreasing $A$, therefore for numerical
calculations, as was already mentioned, we rely upon the exact
Glauber expressions throughout the paper. In what follows we neglect
the real part of the elastic amplitude which gives quite a small
correction, $\sim \rho_{pp}^2/A^{2/3}$, and which otherwise can be
easily implemented.

The simplest process with a Large Rapidity Gap (LRG) is elastic
scattering. It is worth noting, however, that this channel is enhanced by
absorptive corrections, while other LRG processes considered below are
suppressed by these corrections.

The elastic cross section according to (\ref{30}) reads,
 \beq
\sela=\int d^2b\, \left|1-
\exp\left[-{1\over2}\,
\st\,\widetilde T^h_A(b)\right]\right|^2\ ,
\label{60}
 \eeq
 where $\widetilde T^h_A(b)$ is given by (\ref{54}).

The quasielastic cross section also gets modifications compared to the Glauber expression Eq.~(\ref{90}).
The nucleon correlations show up in the second order in nuclear density, leading
to an additional term proportional to $(\sigma^{pN}_{in})^2\,\Delta T^h_A(b)$.
Thus, the cross section of quasielastic proton-nucleus scattering, $pA\to pA^*$, gets the form,
 \beqn
&& \sigma^{pA}_{qel}=
\int d^2b\,\Biggl\{
\exp\left[-\sinhad\,T^h_A(b) -
\frac{(\sigma^{pN}_{in})^2}{\st}\,\Delta T^h_A(b)\right]
\nonumber \\ &-&
\exp\left[-\st\Bigl(T^h_A(b)+\Delta T^h_A(b)\Bigr)\right]\Biggr\}.
\label{90a}
 \eeqn
Notice that in deriving this expression we implicitly used the
assumption that the impact parameter dependence of powers of the
amplitude $\Gamma^{pN}(s)$ does not depend on the power. Although
this is certainly not correct, the approximation is rather accurate
as far as the $NN$ interaction radius is much smaller than the size
of the nucleus. Nevertheless, we used this approximation only for
the sake of clarity and simplicity. For numerical calculations, we
use the more complicated but exact analogue of Eq.~(\ref{90a}).

\section{Gribov corrections via light-cone dipoles}\label{lc}

The dipole representation for the amplitude of hadronic interactions
allows to sum up the Gribov inelastic corrections to all order. We
assume the collision energy to be high enough to keep the dipole
size "frozen" by Lorentz time delation during propagation through
the nucleus. In this limit the calculations are much simplified.

The key ingredients of the approach are the universal dipole-nucleon
cross section and the light-cone wave function of the projectile
hadron \cite{zkl}. Several different models were tested in
\cite{kps}, by comparing with data on proton diffraction. Here we
select two models which describe diffraction quite well. Both employ
the saturated shape of the dipole cross section and differ only by
modeling the proton wave function.

In the limit of soft interactions the Bjorken $x$ is not a proper variable
any more, and the dipole cross section should depend on energy. We rely on
the model proposed in \cite{kst2} and fitted to data,
 \beq
\sigma_{\bar qq}(r_T,s)=\sigma_0(s)\,\left[
1-{\rm exp}\left(-\frac{r_T^2}
{R_0^2(s)}\right)\right]\ ,
\label{180}
 \eeq
 where $R_0(s)=0.88\,fm\,(s_0/s)^{0.14}$ and $s_0=1000\,GeV^2$
\cite{kst2}. The energy dependent factor $\sigma_0(s)$ is defined as,
 \beq
\sigma_0(s)=\sigma^{\pi p}_{tot}(s)\,
\left(1 + \frac{3\,R^2_0(s)}{8\,\la r^2_{ch}\ra_{\pi}}
\right)\ ,
\label{190}
 \eeq
 where $\la r^2_{ch}\ra_{\pi}=0.44\pm 0.01\,fm^2$ \cite{pion} is the mean
square of the pion charge radius.
This dipole cross section is normalized to reproduce the pion-proton total
cross section, $\la\sigma_{\bar qq}\ra_\pi=\sigma_{tot}^{\pi p}(s)$.

For the proton wave function we employ two models.

\subsection{$q-2q$ model}

There are many evidences (although neither of them looks decisive) for a strong paring of the $u$ and $d$ valence quarks into a small size scalar-isoscalar diquark \cite{sb,diquark,kz}.
Neglecting the diquark radius
we arrive at a meson-type color dipole structure of the proton,
 \beq
\left|\Psi_N(\vec r_1,\vec r_2,\vec r_3)\right|^2 =
\frac{2}{\pi\,R_p^2}
\exp\left(-\frac{2\,r_T^2}{R_p^2}\right)\ ,
\label{210}
 \eeq
 where $\vec r_i$ are the interquark transverse distances, $\vec r_3=0$,
 $\vec r_T=\vec r_1=\vec r_2$, and $R_p$ is related to
the mean charge radius squared of the proton as $R_p^2={16\over 3}\la
r^2_{ch}\ra_p$. The dipole wave function squared, Eq.~(\ref{210}), convoluted with the dipole cross section, Eq.~(\ref{180}), gives the proton-proton total cross section.

In this model the effect of the Gribov corrections in all orders is
equivalent to the replacement of the Glauber formula Eq.~(\ref{50})
by,
 \beqn
\sta &=&
2\int d^2b\int\limits_0^1 d\alpha\int d^2r_T\,
|\Psi_N(r_T,\alpha)|^2
\nonumber\\ &\times&
\left[1-e^{-\frac{1}{2}\sigma_{\bar qq}(r_T,s)
T^{\bar qq}_A(b,r_T,\alpha)}\right]
\nonumber\\ &\equiv& 2\int d^2b
\left[1-\left\la e^{-\frac{1}{2}\sigma_{\bar qq}(r_T,s)
T^{\bar qq}_A(b,r_T,\alpha)}\right\ra\right].
\label{212}
 \eeqn
Here we consider a $\bar qq$ (or $qq-q$) dipole of transverse
separation $\vec r_T$ and fractional light-cone momenta $\alpha$ and
$1-\alpha$ of the constituents. The integration over these variables
weighted by the hadron wave function squared is denoted as
averaging. The new notation $T^{\bar qq}_A(b,r_T,\alpha)$ is, \beq
T^{\bar qq}_A(b,r_T,\alpha)=\frac{2}{\sigma_{\bar qq}(r_T)}\int d^2l
\Re \Gamma^{\bar qqN}(\vec l,\vec r_T,\alpha)T_A(\vec b-\vec l).
\label{214} \eeq The partial dipole-nucleon elastic amplitude $\Re
\Gamma^{\bar qqN}(\vec l,r_T,\alpha)$, corresponding to the dipole
cross section (\ref{180}), was derived recently in
\cite{partial1,partial2,partial3},
 \beqn
&&\Re \Gamma^{\bar qqN}(\vec l,\vec r_T,\alpha) =
\frac{\sigma_0(s)}{8\pi B(s)}
\nonumber \\ &\times&
\Biggl\{\exp\left[-\frac{[\vec l+\vec r_T(1-\alpha)]^2}{2B(s)}\right] +
\exp\left[-\frac{(\vec l-\vec r_T\alpha)^2}{2B(s)}\right]
\nonumber \\ &-&
2\exp\Biggl[-\frac{r_T^2}{R_0^2(s)}
-\frac{[\vec l+(1/2-\alpha)\vec r_T]^2}{2B(s)}\Biggr]
\Biggr\},
\label{216}
 \eeqn
where $B(s)=B^{pN}_{el}(s)-{1\over3}\la r_{ch}^2\ra_p-
{1\over8}R_0^2(s)$. It is easy to check that this partial amplitude correctly reproduces the dipole-nucleon cross section Eq.~(\ref{180}),
\beq
\sigma_{\bar qq}(r_T,s)=2\int d^2l\,
\Re \Gamma^{\bar qqN}(\vec l,\vec r_T,\alpha),
\label{218}
\eeq
and the slope of the differential elastic $pN$ scattering,
\beq
B^{pN}_{el}(s)=\frac{1}{\sigma^{pN}_{tot}}
\int d^2l\,\,l^2
\left\la \Re \Gamma^{\bar qqN}(\vec l,\vec r_T,\alpha)\right\ra.
\label{219}
\eeq

These properties of the partial amplitude lead to the following
relations with the analogous functions defined above within the
Glauber model, \beqn \st&=&\Bigl\la \sigma_{\bar qq}(r_T)\Bigr\ra;
\label{221a}\\
\Re\Gamma^{pN}(l)&=&\left\la\Re \Gamma^{\bar qqN}(\vec l,\vec r_T,\alpha)\right\ra;
\label{221b}\\
T^h_A(b)&=&\frac{1}{\st}\Bigl\la \sigma_{\bar qq}(r_T)\,T^{\bar
qq}_A(b,r_T,\alpha)\Bigr\ra. \label{221c} \eeqn Thus, the difference
between the Glauber formula Eq.~(\ref{50}), and the exact
expression, Eq.~(\ref{212}), is in how the averaging over $r_T$ and
$\alpha$ is done: in the former case the averaging is done up in the
exponent, while in the latter case the whole exponential is
averaged.

Notice that $T^{\bar qq}_A(b,r_T,\alpha)$ in the exponent in
Eq.~(\ref{212}) can be replaced by $T^h_A(b)$ with a high precision.
Indeed, at small $r_T^2\ll R_0^2(s)$ the partial amplitude
Eq.~(\ref{216}) vanishes as $\Re \Gamma^{\bar qqN}(\vec l,\vec
r_T,\alpha) \propto r_T^2$. This $r_T$ dependence cancels in
(\ref{214}) with the same behavior of $\sigma_{\bar qq}(r_T)$ in the
denominator. Thus, $T^{\bar qq}_A(b,r_T,\alpha)$ is independent of
$r_T$ in this limit. In the opposite limit of large $r_T^2\gg
R_0^2(s)$ the last term in (\ref{212}) vanishes, and the amplitude
integrated over $d^2l$ becomes a constant. Moreover, the denominator
of Eq.~(\ref{214}) is independent of $r_T$ in this limit. Thus, one
can neglect the slow $r_T$ dependence of  $T^{\bar
qq}_A(b,r_T,\alpha)$ in Eq.~(\ref{221c}) in comparison with the fast
varying function  $\sigma_{\bar qq}(r_T)$, which is equivalent to
the replacement $T^{\bar qq}_A(b,r_T,\alpha) \Rightarrow T^h_A(b)$.
We rely on this approximation in Eq.~(\ref{212}) and in what
follows.

Thus, for the total $p-A$ cross section we recover the standard expression \cite{zkl,kps},
 \beqn
\sta &=&
2\int d^2b\int d^2r_T\,
|\Psi_N(r_T)|^2
\nonumber\\ &\times&
\Biggl\{1-
\exp\left[-\frac{1}{2}\sigma_{\bar qq}(r_T,s)\,
T^h_A(b)\right]\Biggr\},
\label{212a}
 \eeqn

\subsection{$3q$ model}

Another extreme is to assume no pairing forces and a symmetric
valence quark wave function, \beqn &&\left|\Psi_N(\vec r_1,\vec
r_2,\vec r_3)\right|^2 = \frac{3}{(\pi\,R_p^2)^2}\,\delta(\vec
r_1+\vec r_2+\vec r_3)\ , \nonumber\\ &\times&
\exp\left(-\frac{r_1^2+r_2^2+r_3^2}{R_p^2}\right)\, \label{200}
 \eeqn
  Here the mean interquark separation squared is
$\la \vec r_i^{\,2}\ra={2\over3}R_p^2 = 2\la r_{ch}^{2}\ra_p$.
\begin{figure}[htp]
 \includegraphics[width=9cm]{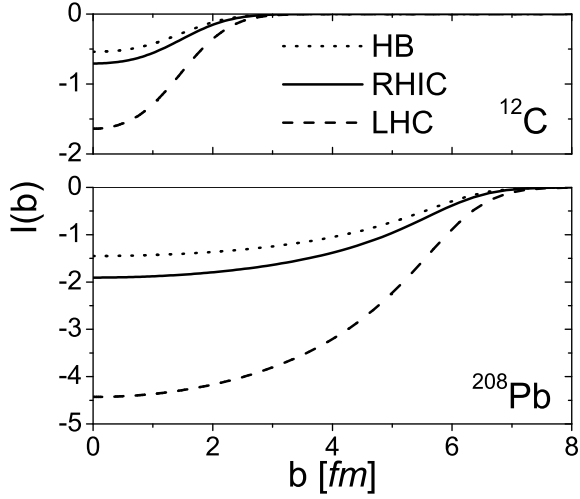}
 \caption{The integral of Eq. (\ref{520}) for carbon and lead,
 calculated at HERA B (\textit{dotted curves}), RHIC (\textit{solid
 curves}) and LHC (\textit{dashed curves}) energies. }
 \label{intB}
 \end{figure}
In this case one needs a cross section for a three-quark dipole,
$\sigma_{3q}(\vec r_1,\vec r_2,\vec r_3)$, where $\vec r_i$ are the
transverse quark separation, with the condition $\vec r_1+\vec
r_2+\vec r_3=0$. In order to avoid the introduction of a new unknown
phenomenological quantity, we express the three-body dipole cross
section via the conventional dipole cross section $\sq$
\cite{mine,kps},
 \beq
\sigma_{3q}(\vec r_1,\vec r_2,\vec r_3) =
{1\over2}\,\Bigl[\sigma_{\bar qq}(r_1)+
\sigma_{\bar qq}(r_2)+
\sigma_{\bar qq}(r_3)\Bigr]\ .
\label{195}
 \eeq
 This form satisfies the limiting conditions, namely, turns into
$\sigma_{\bar qq}(r)$ if one of three separations is zero.

In this model the Gribov corrections modify the Glauber expression Eq.~(\ref{50}) as,
 \beqn
&&\sta =
2\int d^2b\int d^2r_1 d^2r_2 d^2 r_3\,
|\Psi_N(r_i)|^2
\nonumber\\ &\times&
\left\{1-
\exp\left[-\frac{1}{2}\,\sigma_{3q}(r_i,s)\,
T^h_A(b)\right]\right\},
\label{215}
 \eeqn

\section{Gribov corrections to the effect of \boldmath$NN$ correlations}

Nucleon correlations lead to further modifications of the exponent in Eq.~(\ref{212}), which correspond to the replacement $T^{\bar qq}_A(b,r_T,\alpha)\Rightarrow T^{\bar qq}_A(b,r_T,\alpha) + \Delta T^{\bar qq}_A(b)$, where
 \beqn
 && \Delta T_A^{\bar qq}(b,r_T,\alpha)=
  \frac{1}{\sigma_{\bar qq}(r_T)}
  \int d^2l_1\,d^2l_2
  \Delta^\perp_A(\vec l_1,\vec l_2)\nonumber\\ &\times&
  \Re \Gamma^{\bar qq}(\vec b-\vec l_1,r_T,\alpha)
  \Re \Gamma^{\bar qq}(\vec b-\vec l_2,r_T,\alpha),
  \label{500}
  \eeqn
Changing the integration variables $d^2l_1d^2l_2\Rightarrow d^2Ld^2\delta$, where
\beqn
\vec L&=&(\vec l_1+\vec l_2)/2;
\nonumber\\
\vec\delta&=&\vec l_1-\vec l_2; \label{502} \eeqn one has,
correspondingly, $\Delta^\perp_A(\vec l_1,\vec l_2)
\Rightarrow\Delta^\perp_A(\vec L,\vec\delta)$. This function is
rather smooth and varies over distances much longer than the
interaction radius. Therefore, we can take it out of the integral in
Eq.~(\ref{500}), fixing it at $\vec L=\vec b$. Then, using the
partial amplitude Eq.~(\ref{216}) one can perform the integration
over $d^2L$ in (\ref{500}) and then average over $r_T$ and $\alpha$.
The result is, \beqn &&\int d^2L\Bigl\la
 \Re \Gamma^{\bar qq}(\vec l_1,r_T,\alpha)
  \Re \Gamma^{\bar qq}(\vec l_2,r_T,\alpha)
\Bigr\ra =
\nonumber\\ &=&
\left[\sigma^{pN}_{el}+\sigma^{pN}_{sd}\right]
\exp\left[-\frac{\delta^2}{4B(s)+R_0^2(s)/2}\right]
\label{510}
\eeqn
Here $\sigma^{pN}_{sd}$ is the single diffraction cross section,
$pN\to XN$; and the relation \cite{zkl,kps,diffraction}
\beq
\int d^2L\,\left\la
 \left[\Re \Gamma^{\bar qq}(\vec L,r_T,\alpha)\right]^2
 \right\ra = \sigma^{pN}_{el}+\sigma^{pN}_{sd}.
 \label{512}
 \eeq
has been used.

Data show that at high energies this cross section is nearly constant and is about $\sigma^{pN}_{sd}\approx 4\mb$, the value which we use in what follows.

Notice that data for single diffraction also include the
contribution from the triple-Pomeron term, which corresponds to
diffractive gluon radiation. This term has not been included so far
in our calculations, which correspond only to diffractive excitation
of the valence quark skeleton of the proton (see in \cite{kps}).
However, the higher Fock components of the light-cone wave function
of the proton should be also added, which effectively incorporate
this contribution by using the total single diffraction cross
section.

Eq.~(\ref{510}) turns into the Glauber model relation Eq.~(\ref{55})
if the diffraction term is removed and the denominator of the
exponent is replaced by $4B_{el}^{pN}$.

Eventually, the correction related to the nucleon correlations to
the nuclear thickness function, convoluted with the dipole cross
section, takes the form, \beqn &&I_A(b)=\Bigl\la \sigma_{\bar
qq}(r_T)\, \Delta T^{\bar qq}_A(b,r_T,\alpha)\Bigr\ra=
\left[\sigma^{pN}_{el}+\sigma^{pN}_{sd}\right] \nonumber\\ &\times&
\int
d^2\delta\,\exp\left[-\frac{\delta^2}{4B(s)+R_0^2(s)/2}\right]\,
\Delta^\perp_A(\vec \delta,b). \label{520} \eeqn
The quantity
$I_A(b)$ is shown in Fig. \ref{intB}, for both $^{12}C$ and
$^{208}Pb$.

Since $\Delta T_A$ is small and its higher orders are negligible,
there is no difference between averaging of the exponential and of
its exponent. Therefore, the result Eq.~{\ref{520}) accounts for all
inelastic shadowing effects in $NN$ correlations. Then, the total
cross section reads,
 \beq
\sta = 2\int d^2b\,
\Biggl\{1-e^{{1\over2}I_A(b)}
\left\la e^{-\frac{1}{2}\sigma_{dip}
T^h_A(b)}\right\ra\Biggr\}.
\label{530}
 \eeq
Here we use a notation which unifies the two models under
consideration. $\sigma_{dip}$ is the dipole cross section, and
averaging corresponds to integration over the light-cone momenta of
the quarks, weighted with the proton wave function squared.

The results of the calculation of the total, elastic and
quasi-elastic cross section for several nuclei and HERA B, RHIC and
LHC energies, obtained within the Glauber approach including NN
correlations, are presented in Tables I, II, III and IV.

\begin{table}[!hp]
  \begin{center}
  \caption{HERA B.}
    \begin{tabular*}{0.48\textwidth}{@{\extracolsep{\fill}}c| c c c c}\hline\hline
$^{12}C$& Glauber & Glauber & q-2q model & 3q model \\
& & +SRC & +SRC & +SRC \\\hline
$\sigma_{tot}$ & 353.71 & 364.11 & 344.16 & 349.37 \\\hline
$\sigma_{el}$  &  86.90 &  92.96 &  82.39 &  85.42 \\\hline
$\sigma_{sd}$  & -      &  -     &   5.43 &   2.40 \\\hline
$\sigma_{sd+g}$& -      &  -     &   0.07 &   0.06 \\\hline
$\sigma_{qe}$  &  22.85 &  19.62 &  21.12 &  22.05 \\\hline
$\sigma_{qsd}$ & -      &  -     &   1.94 &   0.84 \\\hline
$\sigma_{tsd}$ & -      &  -     &  12.47 &  12.92 \\\hline
$\sigma_{dd}$  & -      &  -     &   0.61 &   0.26 \\\hline\hline
$^{27}Al$& & & &\\\hline
$\sigma_{tot}$ & 697.32 & 714.35 & 675.93 & 688.08 \\\hline
$\sigma_{el}$  & 201.02 & 212.26 & 188.22 & 196.28 \\\hline
$\sigma_{sd}$  & -      & -      &  10.82 &   4.56 \\\hline
$\sigma_{sd+g}$& -      & -      &   0.12 &   0.11 \\\hline
$\sigma_{qe}$  &  36.39 &  31.75 &  33.24 &  34.86 \\\hline
$\sigma_{qsd}$ & -      & -      &   2.92 &   1.23 \\\hline
$\sigma_{tsd}$ & -      & -      &  19.47 &  20.42 \\\hline
$\sigma_{dd}$  & -      & -      &   0.91 &   0.38 \\\hline\hline
$^{48}Ti$& & & &\\\hline
$\sigma_{tot}$ & 1113.52 & 1135.53 & 1074.67 & 1095.93 \\\hline
$\sigma_{el}$  &  353.89 &  369.77 &  327.78 &  342.93 \\\hline
$\sigma_{sd}$  & -       & -       &   16.86 &    6.86 \\\hline
$\sigma_{sd+g}$& -       & -       &    0.17 &    0.16 \\\hline
$\sigma_{qe}$  &   48.73 &   42.82 &   44.57 &   46.86 \\\hline
$\sigma_{qsd}$ & -       & -       &    3.85 &    1.61 \\\hline
$\sigma_{tsd}$ & -       & -       &   26.11 &   27.45 \\\hline
$\sigma_{dd}$  & -       & -       &    1.20 &    0.50 \\\hline\hline
    \end{tabular*}
  \end{center}\label{TabHERA1}
\end{table}

In our calculations nuclear
densities which give the correct nuclear rms radius have adopted, and this is a
reason of some differences with the results of Ref.\cite{kps} in the case of
Glauber calculations. The parameters for the total nucleon-nucleon cross section
and the slope of the Glauber profiles have been obtained as in Ref.\cite{kps}

\section{Diffractive excitation of the proton in \boldmath$pA$ collisions}\label{coherent}
While the Glauber model, which is a single channel approximation,
cannot go beyond elastic scattering, the dipole approach treats
diagonal and off-diagonal diffractive channels on the same footing.
Although the calculation of exclusive channels of diffractive
excitation needs knowledge of the light-cone wave function of the
final state (e.g. see \cite{photo1,photo2}), the total cross section
of diffractive excitation summed over final states is easier to
obtain, since one can employ completeness.
\begin{table}[!hp]
  \begin{center}
  \caption{HERA B (continuation of Table I)}
    \begin{tabular*}{0.48\textwidth}{@{\extracolsep{\fill}}c| c c c c}\hline\hline
$^{184}W$  & Glauber & Glauber & q-2q model & 3q model \\
& & +SRC & +SRC & +SRC \\\hline
$\sigma_{tot}$ & 2972.02 & 2986.46 & 2688.09 & 2747.66 \\\hline
$\sigma_{el}$  & 1174.09 & 1187.18 & 1025.16 & 1074.48 \\\hline
$\sigma_{sd}$  & -       & -       &   32.27 &   11.75 \\\hline
$\sigma_{sd+g}$& -       & -       &    0.28 &    0.23 \\\hline
$\sigma_{qe}$  &   67.04 &   58.35 &   57.60 &   60.89 \\\hline
$\sigma_{qsd}$ & -       & -       &    4.92 &    2.04 \\\hline
$\sigma_{tsd}$ & -       & -       &   33.74 &   35.66 \\\hline
$\sigma_{dd}$  & -       & -       &    1.54 &    0.64 \\\hline\hline
$^{197}Au$ & & & &\\\hline
$\sigma_{tot}$ & 2976.26 & 2989.94 & 2859.84 & 2920.75 \\\hline
$\sigma_{el}$  & 1193.54 & 1206.10 & 1100.54 & 1150.99 \\\hline
$\sigma_{sd}$  & -       & -       &   32.55 &   11.95 \\\hline
$\sigma_{sd+g}$& -       & -       &    0.29 &    0.24 \\\hline
$\sigma_{qe}$  &   62.94 &   54.69 &   61.15 &   64.53 \\\hline
$\sigma_{qsd}$ & -       & -       &    5.31 &    2.22 \\\hline
$\sigma_{tsd}$ & -       & -       &   35.82 &   37.80 \\\hline
$\sigma_{dd}$  & -       & -       &    1.66 &    0.69 \\\hline\hline
$^{208}Pb$ & & & &\\\hline
$\sigma_{tot}$ & 3052.11 & 3117.62 & 2955.57 & 3018.21 \\\hline
$\sigma_{el}$  & 1243.00 & 1274.60 & 1147.01 & 1199.14 \\\hline
$\sigma_{sd}$  & -       & -       &   32.88 &   12.02 \\\hline
$\sigma_{sd+g}$& -       & -       &    0.29 &    0.24 \\\hline
$\sigma_{qe}$  &   62.55 &   54.11 &   61.01 &   64.39 \\\hline
$\sigma_{qsd}$ & -       & -       &    5.31 &    2.21 \\\hline
$\sigma_{tsd}$ & -       & -       &   35.73 &   37.71 \\\hline
$\sigma_{dd}$  & -       & -       &    1.66 &    0.69 \\\hline\hline
    \end{tabular*}
  \end{center}\label{TabHERA2}
\end{table}
Following the standard classification of diffractive channels in terms of
the triple Regge approach \cite{kklp}, one can consider diffractive
excitation of the valence quark system, which corresponds to
the Pomeron-Pomeron-Reggeon ($\Pom\Pom\Reg$) term, and diffractive gluon
radiation corresponding to the triple Pomeron term ($\Pom\Pom\Pom$). The
former mostly contributes
to small mass excitations, $d\sigma/dM_X^2\propto 1/M_X^3$, while the
latter is responsible for the large mass tail,
$d\sigma/dM_X^2\propto 1/M_X^2$, where $M_X$ is the invariant mass of the
produced system, $pp\to Xp$.

\subsection{Coherent excitation of the projectile valence quark system}\label{beam dd}
The cross section of coherent single diffraction on a nucleus,
caused by excitation of the valence quark skeleton without gluon
radiation, is given as usual by the dispersion of the distribution
of eigen elastic amplitudes, where the eigenstates are the dipoles
\cite{zkl,diffraction}.
 \beqn
&&
\sigma_{sd}(pA\to XA)_{\Pom\Pom\Reg} =
\int d^2b\,e^{I_A(b)}
\nonumber\\ &\times&
\left[
\left\la e^{-\sigma_{dip}\, T^h_A(b)}\right\ra -
\left\la e^{-{1\over2}\,
\sigma_{dip}\, T^h_A(b)}\right\ra^2\right],
\label{318}
 \eeqn
 where $I_A(b)$ is given by Eq.~(\ref{520}).
 Dependent on the model, the dipole cross section here has the form of either Eq.~(\ref{180}), or (\ref{195}), and the averaging is weighed by the wave function squared having the form of either Eq.~(\ref{210}), or (\ref{200}).
\begin{table}[!htp]
  \begin{center}
  \caption{RHIC.}
    \begin{tabular*}{0.48\textwidth}{@{\extracolsep{\fill}}c| c c c c}\hline\hline
$^{12}C$& Glauber & Glauber& q-2q model & 3q model \\
& & +SRC & +SRC & +SRC \\\hline
$\sigma_{tot}$ & 413.71 & 425.73 & 406.90 & 410.20 \\\hline
$\sigma_{el}$  & 112.13 & 119.68 & 109.16 & 111.29 \\\hline
$\sigma_{sd}$  & -      & -      &   3.13 &   1.20 \\\hline
$\sigma_{sd+g}$& -      & -      &   0.31 &   0.30 \\\hline
$\sigma_{qe}$  &  26.40 &  23.09 &  26.13 &  26.72 \\\hline
$\sigma_{qsd}$ & -      & -      &   0.95 &   0.29 \\\hline
$\sigma_{tsd}$ & -      & -      &  10.90 &  11.14 \\\hline
$\sigma_{dd}$  & -      & -      &   0.95 &   0.29 \\\hline\hline
$^{208}Pb$& &  &  & \\\hline
$\sigma_{tot}$ & 3297.56 & 3337.57 & 3228.11 & 3262.58 \\\hline
$\sigma_{el}$  & 1368.36 & 1398.08 & 1314.04 & 1343.76 \\\hline
$\sigma_{sd}$  & -       & -       &   16.78 &    5.03 \\\hline
$\sigma_{sd+g}$& -       & -       &    1.06 &    0.98 \\\hline
$\sigma_{qe}$  &   66.06 &   58.47 &   71.99 &   73.92 \\\hline
$\sigma_{qsd}$ & -       & -       &    2.39 &    0.56 \\\hline
$\sigma_{tsd}$ & -       & -       &   30.03 &   30.83 \\\hline
$\sigma_{dd}$  & -       & -       &    2.39 &    0.56 \\\hline\hline
    \end{tabular*}
  \end{center}\label{TabRHIC}
\end{table}

Although Gribov corrections to the total cross section are known to
be small, well within $10\%$ \cite{murthy,gsponer}, this is because
they affect only the second exponential term in Eq.~(\ref{50}),
which is small. However, this term itself is modified significantly
by the inelastic shadowing corrections. Therefore, one should expect
a considerable increase of both terms in (\ref{318}) due to
inelastic corrections, which make the nuclear medium considerably
more transparent compared to the Glauber model \cite{zkl}.
Nevertheless, both terms are small for heavy nuclei and suppress
diffraction everywhere except at the nuclear periphery. Thus, the
cross section of single diffraction should rise as $A^{1/3}$, with a
coefficient which is sensitive to the inelastic shadowing
corrections and $NN$ correlations.

The details of the calculations with both models under consideration
can be found in \cite{kps}. The numerical results for several nuclei
and energies are presented in Tables I, II, III and IV.

\begin{table}[!htp]
  \begin{center}
  \caption{LHC.}
    \begin{tabular*}{0.48\textwidth}{@{\extracolsep{\fill}}c| c c c c}\hline\hline
$^{12}C$& Glauber & Glauber & q-2q model & 3q model \\
& & +SRC & +SRC & +SRC \\\hline
$\sigma_{tot}$ & 598.79 & 613.68 & 591.05 & 592.12 \\\hline
$\sigma_{el}$  & 198.11 & 208-59 & 194.84 & 195.65 \\\hline
$\sigma_{sd}$  & -      &  -     &   0.74 &   0.20 \\\hline
$\sigma_{sd+g}$& -      &  -     &   2.58 &   2.56 \\\hline
$\sigma_{qe}$  &  49.10 &  45.42 &  45.03 &  45.22 \\\hline
$\sigma_{qsd}$ & -      &  -     &  -0.66 &  -0.86 \\\hline
$\sigma_{tsd}$ & -      &  -     &   6.97 &   7.00 \\\hline
$\sigma_{dd}$  & -      &  -     &  -0.66 &  -0.86 \\\hline\hline
$^{208}Pb$ & & & &\\\hline
$\sigma_{tot}$ & 3850.63 & 3885.77 & 3833.26 & 3839.26 \\\hline
$\sigma_{el}$  & 1664.76 & 1690.48 & 1655.70 & 1660.67 \\\hline
$\sigma_{sd}$  & -       & -       &    2.62 &    0.59 \\\hline
$\sigma_{sd+g}$& -       & -       &    2.58 &    2.56 \\\hline
$\sigma_{qe}$  &  120.92 &  112.65 &  113.37 &  113.88 \\\hline
$\sigma_{qsd}$ & -       & -       &   -2.08 &   -2.62 \\\hline
$\sigma_{tsd}$ & -       & -       &   17.55 &   17.63 \\\hline
$\sigma_{dd}$  & -       & -       &   -2.08 &   -2.62 \\\hline\hline
    \end{tabular*}
  \end{center}\label{TabLHC}
\end{table}

\subsection{Coherent diffractive gluon radiation}
\label{gluons}

Diffractive gluon radiation also contributes to the single diffractive
process $pA\to XA$.  Correspondingly, the single-diffraction cross
section Eq.~(\ref{318}) must be corrected for this excitation channel. The
cross section of coherent gluon radiation on a nucleus is given by \cite{kst2},
 \beqn
&&\sigma_{sd}(pA\to XA)_{3\Pom}=
\frac{3}{4\pi}\,\ln\left[\frac{s(1-x_0)}{M_0^2}\right]
\nonumber\\ &\times&
\int d^2b\,e^{I_A(b)}\,
\left\la e^{-{1\over2}\,
\sigma_{dip}\, T^h_A}\right\ra^2
\int d^2r_T\,\biggl|\Psi_{qG}(\vec r_T)\biggr|^2
\nonumber\\ &\times&
\left\{1-\exp\left[-{9\over16}\left(
\sigma_{\bar qq}(r_T,s) T^h_A(b)-{9\over8}I_A(b)\right)\right]\right\}^2.
\label{390}
 \eeqn
Here $M_0^2=5\GeV^2$ is the minimal effective mass squared of the
proton excitation, $x_0=0.85$ is the minimal value of Feynman $x$,
which can be treated as being in the domain of the  triple-Regge
kinematics. \cite{kklp}.

The first factor in (\ref{390}) accounts for the absorptive
corrections, which are due to the lack of initial/final state
interaction of the valence quarks propagating through the nucleus.
Further details about the calculations can be found in \cite{kps}.
The numerical results for several nuclei and energies are presented
in Tables I, II, III and IV.

\section{Quasielastic scattering with and without excitation of the
projectile}\label{incoherent}

In the cases when either the beam proton ($pA\to XA$), or the
nucleus ($pA\to pA^*$), or both ($pA\to XA^*$), are diffractively
excited, one can make use of completeness, which substantially
simplifies the calculations. As was already mentioned, the important
condition for the nucleus is that it decays into nuclear fragments
with no new particle produced. The dipole formalism for these
processes was developed in \cite{kps}. Here we rely on those results
and introduce corrections related to $NN$ correlations.

The simplest processes are double excitation $pA\to XA^*$, where $X$
includes the ground state proton, as well as quasi-diffraction, with
all diffractive excitation of its valence quark system (without
gluon radiation), and breakup of the nucleus. All channels of
coherent interactions which leave the nucleus intact, should be
subtracted,
 \beqn
&& \sigma_{qel}(pA\to pA^*) +
\sigma_{qsd}(pA\to XA^*)
\nonumber\\ &=&
\int d^2b\, \Biggl\la
e^{- \sigma_{dip}\T}
\left\{e^{\tilde I_A(b)}
e^{\frac{\sigma_{dip}^2\T}
{16\pi B_{el}}} -e^{I_A(b)}\right\}
\Biggr\ra
\nonumber\\ &=&
\int d^2b\,
\sum\limits_{k=0}\,\frac{1}{k!}\,
\left[\frac{\T}{16\pi B_{el}}\right]^k
\left[e^{\tilde I_A(b)}-e^{I_A(b)}\delta_{k0}\right]
\nonumber\\ &\times&
\frac{\partial^{2k}}{\partial(T^h_A)^{2k}}\,
\left\la
e^{- \sigma_{dip}\,\T}\right\ra.
\label{600}
 \eeqn
Here besides the function $I_A(b)$ defined in (\ref{520}) we
introduce a new one, \beqn &&\tilde I_A(b)= \int
d^2l_1\,d^2l_2\,\Delta^\perp_A(\vec l_1,\vec l_2) \nonumber\\
&\times& \Biggl\la \left[2\Re \Gamma^{\bar qq}(\vec l_1,\vec
r_T,\alpha)- \left(\Re \Gamma^{\bar qq}(\vec l_1,\vec
r_T,\alpha)\right)^2\right] \nonumber\\ &\times& \left[2\Re
\Gamma^{\bar qq}(\vec l_2,\vec r_T,\alpha)- \left(\Re \Gamma^{\bar
qq}(\vec l_2,\vec r_T,\alpha)\right)^2\right] \Biggr\ra \nonumber\\
&\approx& \left[\frac{\st-\sel-\sigma_{sd}^{pN}}{\st}\right]^2
I_A(b) \label{620} \eeqn
In order to simplify the calculations, we
neglect here the difference in the slopes of powers of the partial
amplitude. This is a second order correction, i.e. a correction to a
correction.

One can single out in (\ref{600}) the quasielastic channel. For that
purpose one should average over the dipole sizes, separately for
both the incoming and outgoing protons,
 \beqn
\sigma^{pA}_{qel} &=&
\int d^2b \Bigl\la\Bigl\la e^{
- {1\over2}\sigma^{(1)}_{dip}\,\T} \,
e^{- {1\over2}\sigma^{(2)}_{dip}\,\T}
\nonumber\\ &\times&
\left.\left.
\left[e^{\tilde I_A(b)}
e^{\frac{1}{16\pi B_{el}}\,\sigma^{(1)}_{dip}\sigma^{(2)}_{dip}\,\T}\ -
e^{I_A(b)}\right]
\right\ra_1
\right\ra_2
\nonumber\\ &=&
\int d^2b\,
\sum\limits_{k=0}\,\frac{1}{k!}\,
\left[\frac{\T}{4\pi B_{el}}\right]^k
\left[e^{\tilde I_A(b)}-e^{I_A(b)}\delta_{k0}\right]
\nonumber\\ &\times&
\left\{\frac{\partial^k}{\partial(T^h_A)^k}\,
\left\la
e^{- {1\over2}\,\sigma_{dip}\,\T}
\right\ra\right\}^2\ .
\label{640}
 \eeqn
 This is a fast converging series due to the smallness of the elastic cross
section. We control the accuracy to be within $1\%$.

Subtracting (\ref{640}) from (\ref{600}) one can get the
quasi-diffractive cross section, which includes the proton
excitations without gluon radiation. To include gluon radiation we
use the same prescription as in (\ref{512}), replacing the
$\Pom\Pom\Reg$ term, $\left[\sigma^{pp}_{sd}\right]_{\Pom\Pom\Reg}$,
by the total single-diffraction cross section.

In the case of nuclear breakup the recoil bound nucleon can be also
diffractively excited. We relate the cross sections for such
channels to the above calculated quasi-elastic and quasi-diffractive
processes, in the same way as in \cite{kps}.

\section{Gluon shadowing}\label{gluon-shad}

In terms of the parton model, gluon shadowing is interpreted in the
nuclear infinite momentum frame as a result of fusion of gluons
originating from different bound nucleons. This process leads to a
reduction of the gluon density in the nucleus at small $x$. The
ultimate form of gluon shadowing is gluon saturation \cite{glr}.

In terms of the dipole approach gluon shadowing is described as
Glauber shadowing for higher Fock states containing gluons,
\cite{kst2}. The effect turns out to be rather weak due to the
shortness of the quark-gluon and gluon-gluon correlation radius, an
observation which is supported by many experimental evidences
\cite{qm04,spots}. For this reason we neglect the small effects of
nucleon correlations in the calculation of gluon shadowing, and use
the results of Ref.~\cite{kps}.

\section{Conclusions}\label{conclusions}

In this paper we further developed the dipole approach of
\cite{zkl,mine,kps} to the calculation of Gribov inelastic
corrections. We employed two models for the proton wave function,
which result in reasonable diffractive cross sections for $pp$
collisions. Here we increased the accuracy of the calculation of the
cross sections of different diffractive processes on nuclei by
improving the model for the nuclear wave function. Namely, we went
beyond the popular single particle density approximation and
introduced corrections for nucleon-nucleon correlations, which lead
to sizable effects, modifying the effective nuclear thickness
function \cite{totalnA}. While inelastic shadowing corrections make
the nuclear medium more transparent for colorless hadrons, the
nucleon short range correlations work in the opposite direction
making the medium more opaque. The influence of both effects on
different diffractive channels vary. They are especially large for
quasielastic and quasi-single diffractive processes associated with
the survival probability of colorless hadrons propagating through a
nuclear medium.

\begin{acknowledgments}

This work was supported in part by Fondecyt (Chile)
grants 1090236 and 1090291, and by DFG (Germany) grant PI182/3-1.
C.d.A. thanks HELEN project for support during his
visit to the Department of Physics,
UTFSM, Valparaiso, where  this work has  been initiated.
M.A. was supported by a DOE grant under contract
DE-FG02-93ER40771, he also thanks HPC-EUROPA2 (project number: 228398)
with the support of the EU - Research Infrastructure Action FP7.
This work was finished during an extended visit
by the authors at INT, University of Washington; they
are thankful to the INT staff for support and warm hospitality.
\end{acknowledgments}


\end{document}